\documentclass[10pt]{article}
\def\squareforqed{\hbox{\rlap{$\sqcap$}$\sqcup$}}
\def\qed{\ifmmode\squareforqed\else{\unskip\nobreak\hfil
\penalty50\hskip1em\null\nobreak\hfil\squareforqed
\parfillskip=0pt\finalhyphendemerits=0\endgraf}\fi} 
\def\bbbn{{\rm I\!N}}
\def\dim{{\rm dim}\,}
\def\bbbr{{\rm I\!R}} %reelle Zahlen 
\def\bbbz{{\mathchoice {\hbox{$\sf\textstyle Z\kern-0.4em Z$}}
{\hbox{$\sf\textstyle Z\kern-0.4em Z$}} 
{\hbox{$\sf\scriptstyle Z\kern-0.3em Z$}} {\hbox{$\sf\scriptscriptstyle
Z\kern-0.2em Z$}}}} 
\def\bbi{{\mathchoice {\hbox{$\sf\textstyle 1\kern-0.25em {\rm I}$}}
{\hbox{$\sf\textstyle 1\kern-0.4em I$}} 
{\hbox{$\sf\scriptstyle 1\kern-0.3em I$}} {\hbox{$\sf\scriptscriptstyle
\kern-0.2em I$}}}} 
\def\bbq{{\mathchoice {\hbox{$\sf\textstyle I\kern-0.4em Q$}}
{\hbox{$\sf\textstyle I\kern-0.4em Q$}} 
{\hbox{$\sf\scriptstyle I\kern-0.3em Q$}} {\hbox{$\sf\scriptscriptstyle
1\kern-0.2em I$}}}} 

\let\mcd=\mathchardef
\mcd\Re="7C52
\mcd\Ze="7C5A
\mcd\Ne="7C4E
\mcd\Le="7C4C
\mcd\Ee="7C45

\newtheorem{theorem}{Theorem}

\newtheorem{definition}{Definition}
\newlength{\au}
\begin{document}
\title{Hausdorff dimension and hierarchical system dynamics}
\author{
\begin{minipage}{\au}K.~Lukierska-Walasek\footnote{e-mail: klukie@proton.if.uz.zgora.pl}\\ 
{\normalsize Institute of Physics\\[-1mm]
University of Zielona G\'ora\\[-1mm] 
ul. Z. Szafrana 4a\\[-1mm]
65-516~Zielona~G\'ora,~Poland\\[-1mm]}
\end{minipage}\and 
\begin{minipage}{\au} K.~Topolski\footnote{e-mail: topolski@math.uni.wroc.pl}\\
{\normalsize Institute of Mathematics,\\[-1mm] 
Wroc\l aw University,\\[-1mm]
Pl.~Grunwaldzki 2/4,\\[-1mm]
50-384 Wroc\l aw, Poland\\[-1mm]}
\end{minipage}}
\date{}
\maketitle
\begin{abstract}

We show that Hausdorff dimension may be used to distinguish different dynamics of the  
relaxation in hierarchical systems.  We examine 
the hierarchical systems following the 
temperature-dependent power-law decay and 
the Kohlrausch law. For our purposes, we consider a L\'evy stochastic processes on $p-$adic
integer numbers.  
\end{abstract}

%\pacs{PACS }
%\keywords{Spin glass, $p-$adic, Markov processes}

\section{\label{sec:level1}Introduction}
The L\'evy stochastic processes are natural generalization of the Brownian motion  \cite{1,2}.
The foundation for this generalization is the theory of the infinitely divisible probability distributions established by
by the works of Khintchine \cite{K, K1} and L\'evy  \cite{1}. The further development is presented in the book of
 Gniedenko and Kolmogorov \cite{KG}.
L\'evy  processes are widely used to describe and model a wide range of physical processes, such as turbulence
\cite{4}, chaotic dynamics \cite{5}, plasma physics \cite{6} and phase transition \cite{DM}. Let us notice that
the L\'evy processes are frequently appear in financial dynamics \cite{8}, biology \cite{9}.

In the case of the Euclidean space the $\alpha-$stable L\'evy processes
are characterized by the L\'evy index $0<\alpha \leq 2$, and when $\alpha =2$
we obtain the Gaussian process. The transition probabilities of the $\alpha$-stable L\'evy process are solution of the
fractional diffusion equation with fractional Riesz derivative \cite{Sa}. The fractal dimension of stable 
L\'evy process with index $\alpha\in (0,1]$ is equal $\alpha$ \cite{BG}.
In the case of the $\bbq_p$ space the similar results for the $\alpha-$stable L\'evy processes hold \cite{Kochubei, Evans}.

In our paper we consider the L\'evy process in the space of $p-$adic numbers.
We consider the Hausdorff dimension of the spherically symmetric L\'evy processes which are
used to describe the relaxation in hierarchical systems. We study the physical system which evolves in time by a
jump from state to state in hierarchical space. Hierarchical structure of metastable states
corresponds to such structure are expressed in terms of pure states in the sense of
 Prisi solution \cite{Mez}.
The transitions between states are thermally activated with rates determined by the free energy barriers separating the
states.
The simplest case is a system with linearly growing, with distance, set of barriers. For such system we have the
temperature-dependent
power law of relaxation dynamics.
Another case is a system  with  barriers grow in slower way, for example proportional to logarithm of distance 
separated states. In this case    the system dynamics fulfills, for large $t,$  the
Kohlrausch law \cite{OS, LT}.\\
To distinguish system with different
dynamics we compare the Hausdorff dimension of the trajectory of the processes describing relaxation.

\section{L\'evy processes with spherically symmetric measure}
Let $p$ be an arbitrary  prime number and let $\bbq_p$ denote the set 
of $p-$adic numbers.
A $p-$adic number is a {\em formal series}, $\sum_{i\geq -M}a_i\,p^i$, 
{\em with coefficients} $a_i$
satisfying
$0\leq a_i \leq p-1$, where $M<\infty$.  
With this definition, a $p-$adic number $a=\sum_{i\geq -M}a_i\,p^i$ 
can be identified with the 
sequence $(a_i)_{i \geq -M}$ of its coefficients.\\
In order to introduce a distance between $p-$adic numbers $a$ and $b$ let us  first 
consider {\em an order} of a $p-$adic number.
The {\em order} of a $p-$adic number $a=(a_i)_{i \geq -M}$ is the smallest $m$
for which $a_m\not =0$
$$
ord_p(a)=\min\{i:\,a_i\not =0\},
$$
with the convention that minimum of the empty set is equal infinity.\\
Now, in term of the function $ord_p(\cdot),$ we may introduce in the  space of $p-$adic 
number the norm $|| \cdot ||_p$  
$$
||x||_p=p^{-ord_p(x)},
$$
and  the $p-$adic metrics 
$$
d_p(x,\,y)=||x-y||_p=p^{-ord_p(x-y)}$$
For  any $x_0\in\bbq _p$ and any integer number $r\in \bbbz $ we may define a closed $p-$adic ball $K(a,p^r)$ 
with center $a$ and radius $p^{r}$
$$K(x_0,p^r)=\left\{x\in \bbq_p,\,\parallel x-x_0\parallel _p\leq p^r\right\}.$$

The family of all such balls constitutes a countable topological base for $\bbq_p$.
By $dx$ we denote the normalized Haar measure, $\int_{\parallel x \parallel _p\leq 1}\,dx=1$,
on Borel $\sigma-$field ${\cal B}$ of subsets of the space $\bbq_p.$\\
Let $\chi(a)$ denotes  a normalized additive character of $\bbq_p$, defined by the formula
$$\chi(a)=\exp(2\pi i \{a\})$$
 where $\{a\}$ is the fractional part of $a\in\bbq_p$.\\

The Fourier transform of a complex valued function $f\in L_1(\bbq_p)$ is defined for $a\in \bbq_p$ by
$$\hat{f}(a)=\int_{\bbq_p}\chi(ax)F(x)dx,$$
%\newpage
%\noindent
We consider the L\'evy process on ${\bbq}_p$, defined as in \cite{Evans} 
\begin{definition}
The Hunt process $X=\left(\Omega,\,{\cal F},\, {\cal F}_t,\,X_t,\,\theta_t,\,P_x\right)$ with state space $\bbq_p$ and
adjoined terminal state $\Delta$ is a L\'evy process on $\bbq_p$ if
\begin{description}
\item[{\rm (1)}] for $t\geq 0$ and $A$ a Borel subset of $\bbq_p$ $P_x\left(X_t\in A\right)=P_0\left(X_t+x\in A\right)$,
\item[{\rm (2)}] for all $t\geq 0$ $P_0(X_t\in\bbq_p)=1 $.
\end{description}
\end{definition}
Let $X$ be a L\'evy process on $\bbq_p$ with transition function $F_t(dx)$. 
The Fourier transform of the transition function of the process  $X$ 
$$\hat{\Phi}(t,a)=\int_{\bbq_p}\chi (xa)F_t(dx),$$
has 
the following L\'evy-Khinchine representation
$$\hat{\Phi}(t,a)=\exp\left\{t\int_{\bbq_p}(\chi (xa)-1)\nu (dx)\right\},$$
where $t\in \bbbr$ and $\nu $ is L\'evy measure of the L\'evy process which satisfies for all $n\in \bbbn$ condition
$$\nu (\bbq_p\backslash \{\parallel x \parallel _p\leq p^n\})<\infty.$$

We now turn to the consideration of special class of L\'evy processes on $\bbq_p$ defined in the following way.\\
Let $\{a(M),\, M\in \bbbz\}$ be a sequence of real numbers satisfying the following two conditions
\begin{description}
\item[(i)] $a(M)\geq a(M+1)$,
\item[(ii)] $\lim_{n\to\infty}a(n)=0.$
\end{description}
The  $\sigma-$finite measure $\nu $ on $\bbq_p$ defined by
\begin{equation}
\label{equ}
\nu (dx)=\sum_{M=-\infty}^{\infty}\frac{a(M)-a(M+1)}{p^M(p-1)}\bbi _{\{\parallel x\parallel _p=p^{M+1}\}}(x)\,dx
\end{equation}
is a spherically symmetric L\'evy measure.\\
The transition probability of a spherically symmetric L\'evy process can be written explicitly \cite{AZ}.
Denote 
$$
P_{M}=(1-p^{-1})\sum_{i=0}^{\infty}
p^{-i}\exp\{t(1-p)^{-1}(pa(M+i)-a(M+i+1))\},
$$
then the  transition probability $P_t$ of the process with L\'evy measure (\ref{equ})
is given by 
$$P_t(x_0,K(x,p^M))=P_M(t)$$
if $\parallel x_0-x\parallel \leq p^{M}$, and by 
$$
P_t(x_0,\,K(x,\,p^M))=p^{1-m}(p-1)^{-1}\left(P_{M+m}(t)-P_{M+m-1}(t)\right),
$$
if $\parallel x_0-x\parallel=p^{M+m}$, $m\ge 1$.\\ 

The  L\'evy processes with spherically symmetric L\'evy measures are examples of Markov processes with $\bbq_p$ as the
state space considered by Albeverio and Karwowski \cite{31,32}. The spherically symmetric L\'evy  processes 
have been studied by probabilistic approach by many authors  
\cite{Evans, 31, 32, Yasuda, Kaneko}, they were used to study the hierarchical systems dynamics in \cite{34, 35, LT} 
The analytical approach to examine spherically symmetric L\'evy  processes on $\bbq_p$ was presented in 
\cite{33,34, 35, 36, Kochubei}. In this approach authors use
the pseudo-differential operators. The simplest of them, the  fractional differentiation operator 
$D^{\alpha }$ ${\alpha > 0}$, was introduced by 
Vladimirov and is
defined by \cite{29}
$$ D^{\alpha }f(x)=\frac{1-p^{\alpha }}{1-p^{-\alpha -1}}
\int_{\bbq_p}\parallel y\parallel ^{-\alpha -1}_p(f(x-y)-f(x))\,dy,$$
where $f\in D(\bbq_p),$ the space of locally constant function with compact supports.
The transition probabilities of $\alpha-$stable L\'evy process on $\bbq_p$ are solution of the following Cauchy problem
$$\frac{\partial}{\partial t}f(x,t)+a(D^{\alpha}f)(x,t)=0,$$ 
where $f(x,0)=\delta (x)$ and $a>0$ \cite{Kochubei}.

\section{Hausdorff dimension of L\'evy processes on $p-$adic integers.}
In this section we first briefly recall the definition of some basic notions about Hausdorff dimension.
Then we will describe Hausdorff dimension for processes studied in this paper. For the general theory of Hausdorff
measure and Hausdorff dimension we refer to \cite{H1, H2, H3}.
The Hausdorff measure is based on the notation of a {\em covering} of the metric space $E$ by sets of finite diameter. 
A {\em covering} of $E$ is an at most countable collection of sets $E_1,\,E_2,...$ with
$$E\subseteq \bigcup_{i=1}^{\infty} E_i.$$
For every $s\leq 0$ we say that the $s-value$ of the covering is
$\sum_{i=1}^{\infty} |E_i|^s,$
where $|E_i|$ denotes the diameter of the set $E_i.$ 

\noindent
\begin{definition}
For every $\alpha \geq 0$ the $\alpha-$Hausdorff measure of a subset $A \subset E$ of a metric 
space $E$ is defined as
$$ {\cal H}^{\alpha }(A)=\lim_{\delta \downarrow 0}{\cal H}_{\delta }^{\alpha }(A)$$
where
$${\cal H}_{\delta }^{\alpha}(A)=
\{inf\left \{\sum_{i=1}^{\infty}|E_i|^{\alpha }\,\,:(E_i ;\, i\in N)\,\,
 \mbox{is a covering of}\,\, A\,\, \mbox{and}\,\,\, |E_i|<\delta \right\}.$$
\end{definition}
\noindent
It can be shown that ${\cal H}^{\alpha }$ is a measure on Borel subsets of $E$ and if $0\leq s < t<\infty$
 and $A\subset E$ then 
\begin{description}
\item[(1)] ${\cal H}^s(A)<\infty$ implies ${\cal H}^t(A)=0$, 
\item[(2)] ${\cal H}^t(A)>0$ implies ${\cal H}^s(A)=\infty$.
\end{description}

Thus we can defined Hausdorff dimension in the following way.
\begin{definition} The Hausdorff dimension of a subset $A$ $\dim(A)$ is defined as
\begin{eqnarray*}
\dim A&=&\inf \{s:\,\, {\cal H}^s(A)=0 \}=\inf  \{s:\,\, {\cal H}^s(A)<\infty \}\\
&=&\sup \{t:\,\, {\cal H}^t(A)>0 \}=\sup \{t:\,\, {\cal H}^t(A)=\infty \}.
\end{eqnarray*}
\end{definition}

In order to study sample path properties of the  L\'evy processes on the space of $p-$adic integers,
we can use the results by Evans, who investigated L\'evy processes on a locally compact totally
disconnected Abelian topological group. 
As in a case of Euclidean L\'evy process
Hausdorff dimension of the path of $p-$adic L\'evy process, $\{X(t);\,t\geq 0\}$, is characterized in terms
of exit times.\\
Let for $n\geq 0$ $$\tau(n)=\inf\left\{t\geq0:\,X(t)\not\in K(0,p^{-n})\right\}$$ with the convention that 
$\inf \emptyset = \infty,$ and denote
$$q(n;\,N)=P\left\{X(t)\not \in K(0,p^{-n}),\, for\, all\, t\in[\tau(n),\,\tau(N))\right\}.$$

For any Borel set $B\subset \bbbr_{+}$ we define the {\em range} or {\em path} to be the random set generated by $X$
$$X(B)=\{x\in \bbq_p :\, x=X(t)\,\,\mbox{for some}\,\, t\in B\}.$$

The Hausdorff dimension of the sample path of spherically symmetric L\'evy processes on $p-$adic integers is
described in the following result (see  (\cite{Evans} for a proof). 
\begin{theorem} Suppose that the exists an integer 
$N$ such that $\nu \left(\bbbz_p\backslash K(p^{-N}\right)> 0,$ and for all 
$\rho > 0$
$$\lim_{n\to\infty} p^{n \rho}q(n;N)=\infty,$$
then
$$P(\beta' \dim B \leq \dim X(B)\leq \beta'' \dim B \,\,\mbox{for all}\,\, B)=1,$$
where
$$\beta'=\inf\left\{\beta\,:\,\liminf_{n\to\infty}p^{-\beta n}q(n;N)\nu \left(\bbbz_p\backslash K(0,p^{-n})\right)=0 \right\},$$
$$\beta''=\inf\left\{\beta\,:\,\limsup_{n\to\infty}p^{-\beta n}q(n;N)\nu \left(\bbbz_p\backslash K(0,p^{-n})\right)=0 \right\}.$$
\end{theorem}
Let $\{X(t):\,t\geq 0\}$ be the L\'evy processes on $p-$adic integers with L\'evy measure $\nu $ 
such that for all 
$n\geq 1$ and for some $\alpha \in (0,\,1)$
$$\nu \left(\bbbz_p\backslash K(0,p^{-n})\right)=p^{-\alpha n},$$
then it was shown by Evans (\cite{Evans}, page 253) that
$$P\left(\dim X(B)=\alpha\, \dim (B)\,\,\,for\,\, all\,\, B\right)=1.$$
In the similar way we can show that for  a L\'evy processes $\{X(t)\,:\,t\geq 0\},$ on $p-$adic integers  
with L\'evy measure $\nu $ such that for all 
$n\geq 1$ and for some $\alpha > 0$
$$\nu \left(\bbbz_p\backslash K(0,p^{-n})\right)=p^{-\alpha \ln n},$$
we have
$$P\left(\dim X(B)=0\,\,\mbox{\em for all}\,\, B\right)=1.$$

Now we use these results to analyze relaxation dynamics in hierarchical system.
The simplest model of diffusion on hierarchical space is the model proposed by Ogielski and Stein \cite{OS}.
They consider a regular  Cayley tree with $M$  levels and fixed branching ratio $p.$ 
The natural ultrametric 
distance $d(k,\,l)$ 
between leaves $k$ and $l$ , is defined as equal to the height 
$m,\,\,m=0,\,1,\,\dots,\,M$ of their closest common ancestor.
Transitions between states are thermally activated. The height of the energy barriers,
 $\Delta_k\,$ $k=1,\,2,\,\dots,$ which the system  overcomes can be ordered in the 
increasing sequence 
$\Delta_1 < \Delta_2 < \dots < \Delta_k \dots.$   
Now, identifying the 
states $x$ and $y$ 
separated by the energy barrier $\Delta_m$  with leaves $k$ and $l,$ the probability of 
moving from state $x$
 to state $y$ may be defined as equal to transition probability from  leaf  $k$ to
 leaf  $l$ separated by the ultrametric distance
 $m.$ Thus dynamics in the space of states
separated by the energy barrier may be studied in terms of appropriate stochastic process 
involving the end points of Cayley tree, as a space of states.\\
The  rate of transition from a starting site at the Cayley tree to another site lying at an ultrametric distance $k$ is given by
$\exp\{-\Delta_k/T\}$, where $T$ is normalized temperature.   
We may represent the end points of regular Cayley tree with $M$ levels and fixed branching ratio $p$ as
a set of disconnected balls with fixed radius $p^{-M}$ covering space of $p-$adic integer $\bbbz_p$
and define an appropriate  spherically symmetric $\bbbz_p-$valued L\'evy process \cite{LT}.\\

For the sequence of barriers linearly growing with 
distance $\Delta_k=\Delta\,k$ for some positive constant $\Delta$ we consider
 $\bbbz_p-$valued L\'evy process $\{X(t);\,t\geq 0\}$ with a L\'evy measure $\nu(dx)$ such that
$$\nu \left(\bbbz_p\backslash K(0,p^{-n})\right)=p^{-\alpha n},$$
where $\alpha =\Delta (T\ln p)^{-1}$.\\ 
If $\Delta$ and $T$ are such that $\Delta (T\ln p)^{-1}<1$, then
Hausdorff dimension of the path $X([0,1])$ is equal to
$\alpha=\Delta (T\ln p)^{-1}$, with probability one.\\

For the sequence of barriers growing in a slower way, 
$\Delta_k=\Delta\,\ln k$,  we consider
 $\bbbz_p-$valued L\'evy process $\{Y(t);\,t\geq 0\}$ with a L\'evy measure $\nu(dx)$ such that
$$\nu \left(\bbbz_p\backslash K(0,p^{-n})\right)=p^{-\alpha \ln n},$$
where $\alpha =\Delta (T\ln p)^{-1}$.
\\ In this case
Hausdorff dimension of the path $Y([0,1])$ of the process is equal 
$0$, with probability one.

%\newpage

\end{document}